\documentclass{piparticle-final}
\usepackage{graphicx}
\usepackage{amsmath}

\usepackage{cite} % To compress citation ranges
\usepackage{epstopdf}

\begin{document}

\volume{6}               % To be inserted by Editor
\articlenumber{060006}   % To be inserted by Editor
\journalyear{2014}       % To be inserted by Editor
\editor{A. Marti}   % To be inserted by Editor
%\reviewers{}% To be inserted by Editor
\received{20 March 2014}     % To be inserted by Editor
\accepted{7 August 2014}   % To be inserted by Editor
\runningauthor{R. Banerjee \itshape{et al.}}  % To be inserted by Editor
\doi{060006}         % To be inserted by Editor

\title{Influence of surface tension on two fluids shearing instability}

\author{Rahul Banerjee,\cite{inst1}\thanks{E-mail:rbanerjee.math@gmail.com}\hspace{0.5em} S. Kanjilal\cite{inst1}}

\pipabstract{Using extended Layzer's potential flow model, we investigate the
effects of surface tension on the growth of the bubble and spike
in combined Rayleigh-Taylor and Kelvin-Helmholtz instability. The
nonlinear asymptotic  solutions are obtained analytically for the
velocity and curvature of the bubble and spike tip. We find that
the surface tension decreases the velocity but does not affect the
curvature, provided surface tension is greater than a critical
value. For a certain condition, we observe that surface tension
stabilizes the motion. Any perturbation, whatever its
magnitude, results stable with nonlinear oscillations.  The
nonlinear oscillations depend on surface tension and relative
velocity shear of the two fluids.}

\maketitle

\blfootnote{
\begin{theaffiliation}{99}
\institution{inst1} St. Paul's Cathedral Mission College, 33/1, Raja Rammohan Roy, Sarani, 700 009 Kolkata, India.
\end{theaffiliation}
}

%\emph{Keywords}: Shear flow; Gravitational force; Surface tension;Rayleigh-Taylor instability; Kelvin-Helmholtz instability;Bubbles;Spikes\\

\section{Introduction}
When two different density fluids are divided by an interface,
the interface becomes unstable with exponential growth under the
action of a constant acceleration acting in the direction perpendicular to the interface from the heavier to lighter fluid or under the action of relative velocity shear of two
fluids. These two types of instabilities are known as
Rayleigh-Taylor and Kelvin-Helmholtz instabilities, respectively.
Temporal development of the nonlinear structure of the interface
consequent to Rayleigh-Taylor or Kelvin-Helmholtz instability is
currently a topic of interest both from theoretical and experimental
points of view. The nonlinear structure is called a bubble if the
lighter fluid penetrates across the unperturbed interface into the
heavier fluid and it is called a spike if the opposite takes place. The
instabilities arise in connection with a wide range of problems
ranging from direct or indirect laser driven experiments in the
ablation region at compression front during the process of
inertial confinement fusion \cite{br07, bu96} to mixing of
plasmas in space plasma systems, such as boundary of planetary
magnetosphere, solar wind and cluster of galaxies \cite{ha04}.
In  high energy density physics(HEDP), formation of supernova
remnant or formation of astrophysical
jets \cite{dr05,ka01,ry02,sp54,rm00} are also seen in these types of
instabilities. In high energy density plasma experiments using
Omega laser \cite{ha09}, Kelvin-Helmholtz instability growth
has recently been observed .

There are several methods to describe the nonlinear structure of
the interface of two constant density fluids under potential
theory and the associated nonlinear dynamics has been studied by
many authors \cite{dl55,vg02,ss03,qz98}. Layzer \cite{dl55}
described the formation of the structure using an expansion near
the tip of the bubble or the spike up to second order in the
transverse coordinates in two dimensional motion and this approach was
extended in Ref. \cite{rb12} for Kelvin-Helmholtz instability. It
is well known \cite{sc68} that the surface tension reduces the
linear Rayleigh-Taylor Growth rate. The lowering in the growth
rate is seen to increase with increase in the wave number $k$ up
to a critical wave number $k_c=\sqrt{\frac{(\rho_h-\rho_l)g}{T}}$,
where $T$ denotes surface tension, and $\rho_h$ and $\rho_l$ are
the densities of the heavier and lighter fluids, respectively. The same effect has been
described by Mikaelian \cite{mik96} for Rayleigh--Taylor
instability in finite thickness and Sung-Ik Sohn \cite{ss09}
described the effect using the Layzer nonlinear potential model. The nonlinear theory influence of surface tension was elaborately
studied by Pullin \cite{pu82} and Garnier \textit{et al.} \cite{ga03}
using numerical methods.

The present paper addresses to the problem of the time
development of the nonlinear interfacial structure caused by
combined Rayleigh--Taylor and Kelvin--Helmholtz instability in
presence of surface tension. It is shown that the growth rate of
the instabilities is affected by the surface tension. The growth
rate of the tip of the  bubble or spike are significantly reduced
due to the surface tension. We observed an oscillatory
stabilization of the interface for large surface tension. This
oscillation depends on the relative velocity shear also. Section
II deals with the basic hydrodynamical equations together with the
geometry involved. Here we assume that the fluids are inviscid and
the motion is irrotational. The investigation of the nonlinear
aspect of the structure of the two fluids interface is facilitated
by Bernoull's equation together with the pressure balance equation
at the interface. The long time asymptotic behavior of the bubble
and spike tip for combined Rayleigh--Taylor and Kelvin--Helmholtz
instabilities is derived in section III.A and III.B, respectively. We
have also discussed the characteristics of the tip of the bubble
and the spike derived analytically and numerically. Finally, we have
concluded the results in section IV.

\section{Basic mathematical model}

We have considered two incompressible fluids separated by an
interface located at $y=0$ in a two-dimensional $x-y$ plane, where
$x$ axis lying normal to the unperturbed fluid interface. The
fluid with density $\rho_h$ is assumed to overlie the fluid with
density $\rho_l$ and gravity is taken along negative $y$-axis.
 In the following discussion, we shall denote the properties of the
 fluid above the interface by the subscript $h$ and below the
 interface by the subscript $l$.
 After perturbation, the nonlinear interface is assumed
to take up a parabolic shape, given by

\begin{eqnarray} \label{eq:1}
y=\eta(x,t)=\eta_{0}(t)+\eta_{2}(t)(x-\eta_{1}(t))^2
\end{eqnarray}
The perturbed interface forms a bubble or spike according to
$\eta_{0}(t)>0$, $\eta_{2}(t)<0$ or $\eta_{0}(t)<0$,
$\eta_{2}(t)>0$. Functions $\eta_{0}(t)$ and $\eta_{1}(t)$ are
related to the position of the tip of the bubble from the
unperturbed interface, i.e, at time $t$ the position of the bubble
tip is $(\eta_{1}(t),\eta_{0}(t))$ and $\eta_{2}(t)$ is related to
the bubble curvature.

In our previous works \cite{rb12,rb11,mrg12a,mrg12,rb13},
we have considered $\eta_{1}(t)=0$ due to the absence of velocity
shear parallel to the unperturbed interface. However, in presence
of streaming motion of the fluids, the tip of the bubble moves
parallel to unperturbed interface with velocity
$\dot{\eta_{1}}(t)$.

According to the extended  Layzer model \cite{dl55,vg02,rb12,rb11}, the velocity potentials describing the motion for the
upper (heavier) and lower (lighter) fluids are assumed to be given
by

\begin{align}\label{eq:2}
\phi_{h}(x,y,t)&=a_{1}(t)\cos{(k(x-\eta_{1}(t))}e^{-k(y-\eta_{0}(t))} \notag \\ 
&+ a_{2}(t)\sin{(k(x-\eta_{1}(t))}e^{-k(y-\eta_{0}(t))} \notag \\ 
&- x U_{h}
\end{align}

\begin{align}\label{eq:3}
\phi_{l}(x,y,t)&=b_{0}(t)y\notag \\
&+b_{1}(t)\cos{(k(x-\eta_{1}(t))}e^{k(y-\eta_{0}(t))} \notag\\
&+ b_{2}(t)\sin{(k(x-\eta_{1}(t))}e^{k(y-\eta_{0}(t))}\notag\\
&- x U_{l}
\end{align}
where $U_{h}$ and $U_{l}$ are streaming velocities of upper and
lower fluids, respectively, and $k$ is the perturbed wave number.

The evolution of the interface $y=\eta(x,t)$ can be determined by
the kinematical and dynamical boundary conditions. The kinematical
boundary conditions are

\begin{eqnarray}\label{eq:4}
\frac{\partial\eta}{\partial t}-\frac{\partial\eta}{\partial
x}\frac{\partial\phi_{h}}{\partial
x}=-\frac{\partial\phi_{h}}{\partial y}
\end{eqnarray}

\begin{eqnarray}\label{eq:5}
\frac{\partial\eta}{\partial x}(\frac{\partial\phi_{h}}{\partial
x} -\frac{\partial\phi_{l}}{\partial x}) =
\frac{\partial\phi_{h}}{\partial y} -
\frac{\partial\phi_{l}}{\partial y}
\end{eqnarray}
and the dynamical boundary condition (first integral of the
momentum equation) is of the form

\begin{align}\label{eq:6}
-\rho_{h(l)}\frac{\partial \phi_{h(l)}}{\partial t}+
\frac{1}{2}\rho_{h(l)}(\vec{\nabla} \phi_{h(l)})^{2}+\rho_{h(l)} g
y\notag \\
=-p_{h(l)}+f_{h(l)}(t)
\end{align}
The pressure boundary condition at two fluid interface including
 surface tension \cite{ss09, mrg12} is
 
\begin{eqnarray}\label{eq:7}
p_{h}-p_{l}=\frac{T}{R}
\end{eqnarray}
where $T$ is the surface tension and $R$ is the radius of curvature.

 Plugging the  condition (7) at the interface
$y=\eta(x,t)$ in Eq. (6), we obtain the following equation.

\begin{align}\label{eq:8}
\rho_{h}[-\frac{\partial\phi_{h}}{\partial t}+\frac{1}{2}
(\vec{\nabla}\phi_{h})^2]-\rho_{l} [-\frac{\partial
\phi_{l}}{\partial t}+\frac{1}{2}(\vec{\nabla}
\phi_{l})^2]\notag\\
+g(\rho_{h}-\rho_{l})y=-\frac{T}{R}+f_{h}-f_{l}
\end{align}

We have restricted our study near the peak of the perturbed
structure where $|k(x-\eta_{1}(t))|\ll 1$. Thus, we can neglect the
terms of $O(|x-\eta_{1}|^{i})$ $(i\geq3)$ \cite{rb12}. With
this point of view, we have

\begin{align}\label{eq:9}
\frac{1}{R}&=2\eta_2\left(1+4\eta_2^2(x-\eta_1)^2\right)^{-\frac{3}{2}}\notag\\
&\approx
2\eta_2\left(1-6\eta_2^2(x-\eta_1)^2\right)
\end{align}
We substitute all the parameters $\eta$, $\phi_{h}$ and $\phi_{l}$
in the kinematic and dynamic boundary conditions represented by
Eqs. (4), (5), (8) and (9), and equate coefficients of
$(x-\eta_{1})^i$,$(i=0,1,2)$ and neglect terms
$O(|x-\eta_{1}|^{i})$ $(i\geq3)$. This yields the following
equations.

\begin{eqnarray}\label{eq:10}
\frac{d\xi_1}{d \tau}=\xi_4
\end{eqnarray}
\begin{eqnarray}\label{eq:11}
\frac{d\xi_2}{d \tau}=V_{h}-\frac{\xi_5(2\xi_3+1)}{2\xi_3}
\end{eqnarray}
\begin{eqnarray}\label{eq:12}
\frac{d\xi_3}{d \tau}=-\frac{1}{2}(6\xi_3+1)\xi_4
\end{eqnarray}
\begin{eqnarray}\label{eq:13}
\frac{k b_{0}}{\sqrt{kg}}=-\frac{12\xi_3 \xi_4}{6\xi_3-1}
\end{eqnarray}
\begin{eqnarray}\label{eq:14}
\frac{k^2 b_{1}}{\sqrt{kg}}=\frac{6\xi_3+1}{6\xi_3-1} \xi_{4}
\end{eqnarray}
\begin{eqnarray}\label{eq:15}
\frac{k^2
b_{2}}{\sqrt{kg}}=\frac{(2\xi_3+1)\xi_{5}-2\xi_{3}(V_{h}-V_{l})}{2\xi_3-1}
\end{eqnarray}
\begin{align}\label{eq:16}
\frac{d\xi_4}{d\tau}&=\frac{N_{1}(\xi_3,r)}{D_{1}(\xi_3,r)}\frac{\xi_4^2}{(6\xi_3-1)}\notag\\
&+\frac{2(1-r)\xi_3(6\xi_3-1)}{D_{1}(\xi_3,r)}\left(1-12\xi_2^2\frac{k^2}{k_c^2}\right)\notag\\
&+\frac{N_{2}(\xi_3,r)}{D_{1}(\xi_3,r)}\frac{(6\xi_{2}-1)\xi_5^2}{2\xi_{3}(2\xi_3-1)^2}\notag\\
&+\frac{2(4\xi_{3}-1)(6\xi_{3}-1)}{D_{1}(\xi_{3},r)(2\xi_{3}-1)^2}\notag\\
&\times[(V_{h}-V_{l})^2\xi_{3}-(V_{h}-V_{l})(2\xi_{3}+1)\xi_5]
\end{align}
and

\begin{align}\label{eq:17}
\frac{d\xi_5}{d\tau}&=-\frac{(2\xi_{3}-1)r\xi_{4}
\xi_{5}}{2\xi_{3}D_{2}(\xi_3,r)}\notag\\
&+\frac{\xi_{4}(6\xi_{3}+1)}{2D_{2}(\xi_3,r)(6\xi_{3}-1)(2\xi_{3}-1)}\\
&\times[4(V_{h}-V_{l})(4\xi_{3}-1)-
\frac{\xi_{5}}{\xi_{3}}(28\xi_{3}^2-4\xi_{3}-1)]\notag
\end{align}
where $r=\frac{\rho_h}{\rho_l}$; $\xi_1=k\eta_0$; $\xi_2=k\eta_1$;
$\xi_3=\frac{\eta_{2}}{k}$; $\xi_{4}=\frac{k^2a_{1}}{\sqrt{kg}}$;
$\xi_{5}=\frac{k^2a_{2}}{\sqrt{kg}}$; $\tau=t\sqrt{kg}$;
$k_c^2=\frac{(\rho_h-\rho_l)g}{T}$ and
$V_{h(l)}=\frac{kU_{h(l)}}{\sqrt{kg}}$ are corresponding
dimensionless quantities. The function $N_{1,2}(\xi_{3},r)$ and
$D_{1,2}(\xi_{3},r)$ are given by

\begin{align}\label{eq:18}
&N_{1}(\xi_3,r)=36(1-r)\xi_{3}^{2}+12(4+r)\xi_{3}+(7-r); \notag\\
&D_{1}(\xi_3,r)=12(r-1)\xi_{3}^{2}+4(r-1)\xi_{3}-(r+1)
\end{align}
and

\begin{align}\label{eq:19}
&N_{2}(\xi_3,r)=16(1-r)\xi_{3}^{3}+12(1+r)\xi_{3}^2-(1+r); \notag\\
&D_{2}(\xi_3,r)=2(1-r)\xi_{3}+(r+1)
\end{align}
The temporal development of the combined effect of Rayleigh--Taylor
and Kelvin--Helmholtz instability is given by Eqs. (10)--(12), (16)
and (17).

\begin{figure}
%\vspace{-3cm} \vbox{\hskip 1.cm \epsfxsize=15cm \epsfbox{fig1.ps}}
\begin{center}
\includegraphics[width=\columnwidth, trim=0 0cm 0 0, clip]{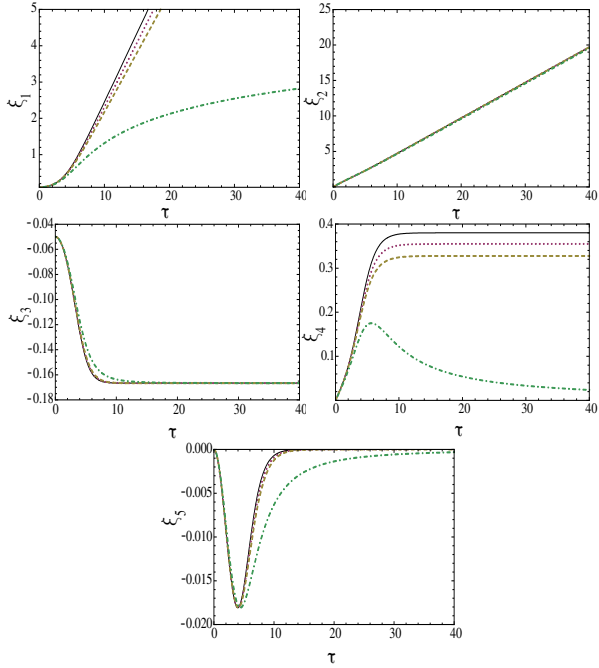}
\end{center}
\caption{Bubble- variation of $\xi_1$, $\xi_2$,
$\xi_3$, $\xi_4$ and $\xi_5$ with $\tau$.  Initial value
$\xi_1=$0.1, $\xi_2=$0, $\xi_3=$-0.05, $\xi_4=0$, and $\xi_5=0$
with $\rho_h=3$, $\rho_l=2$, $V_{h}$=0.5, $V_{l}=0.1$,
$\frac{k^2}{k_c^2}$=0 (line), 0.5 (dot), 1 (dash), 3.9 (dash-dot).} \label{FIG:1}
\end{figure}

\begin{figure}
%\vspace{-5cm} \vbox{ \hskip 1.cm \epsfxsize=15cm \epsfbox{fig2.ps}}
\begin{center}
\includegraphics[width=\columnwidth, trim=0 0cm 0 0, clip]{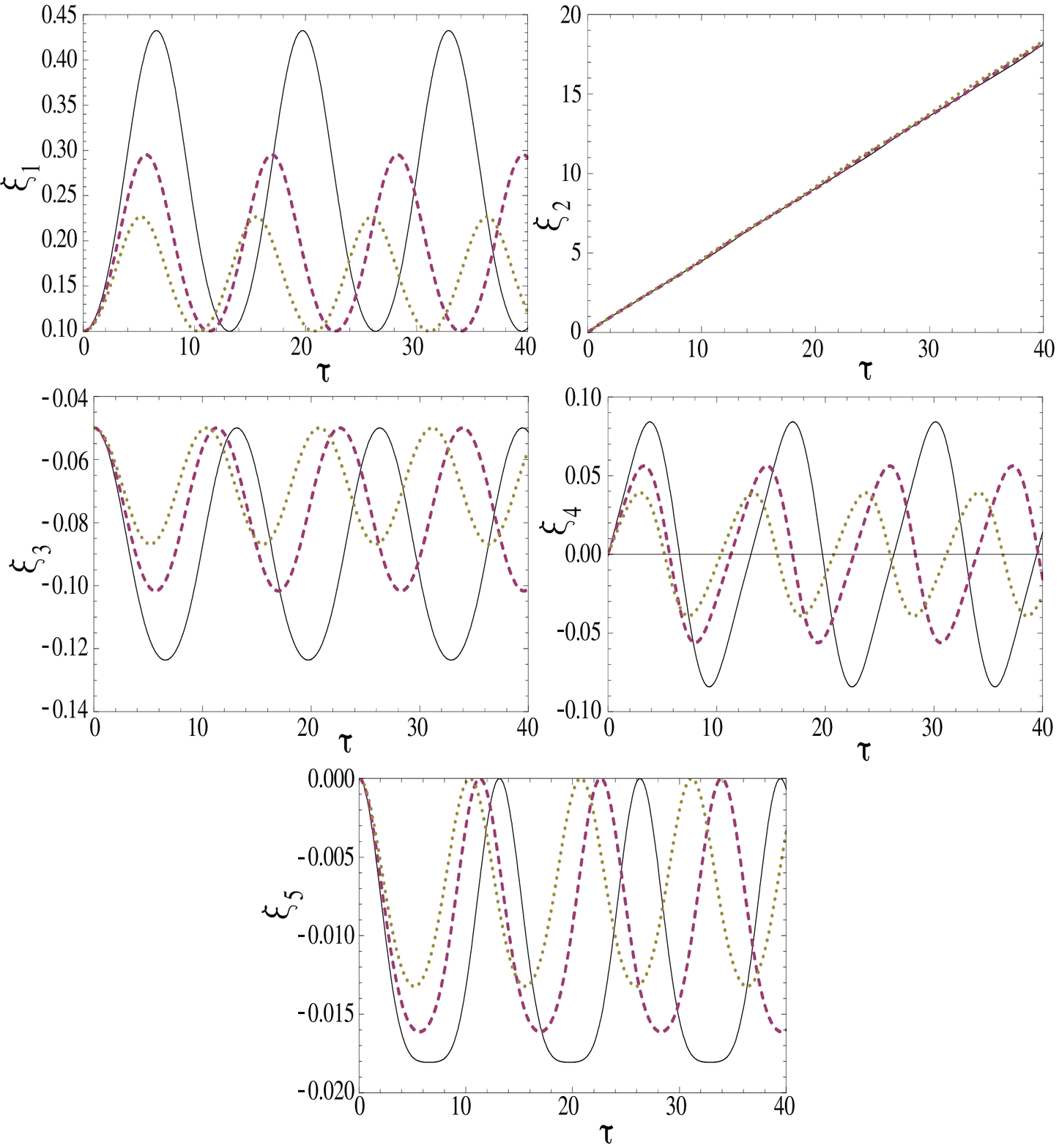}
\end{center}
\caption{Bubble- variation of $\xi_1$, $\xi_2$,
$\xi_3$, $\xi_4$ and $\xi_5$ with $\tau$. Initial value
$\xi_1=$0.1, $\xi_2=$0, $\xi_3=$-0.05, $\xi_4=0$, and $\xi_5=0$
with $\rho_h=3$, $\rho_l=2$, $V_{h}$=0.5, $V_{l}=0.1$,
$\frac{k^2}{k_c^2}$=10 (line), 15 (dot), 20 (dash). } \label{FIG:2}
\end{figure}

\begin{figure}
%\vspace{-2cm} \vbox{ \hskip 1.cm \epsfxsize=15cm\epsfbox{fig3.ps}}
\begin{center}
\includegraphics[width=\columnwidth, trim=0 0cm 0 0, clip]{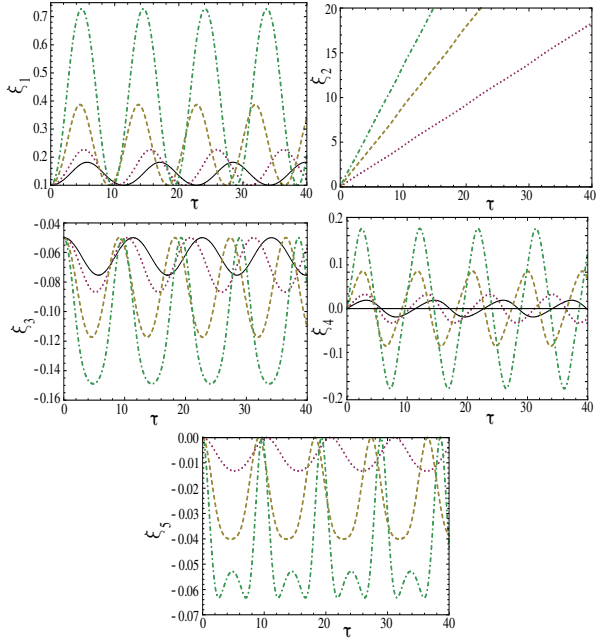} 
\end{center}
\caption{Bubble- variation of $\xi_1$, $\xi_2$,
$\xi_3$, $\xi_4$ and $\xi_5$ with $\tau$. Initial value
$\xi_1=$0.1, $\xi_2=$0, $\xi_3=$-0.05, $\xi_4=0$, and $\xi_5=0$
with $\rho_h=3$, $\rho_l=2$, $\frac{k^2}{k_c^2}$=20,
$V_{h}$=0, $V_{l}=0$ (line), $V_{h}$=0.5, $V_{l}=0.1$(dot),
$V_{h}$=1, $V_{l}=0.1$ (dash), $V_{h}$=1.5, $V_{l}=0.1$ (dash-dot). }
\label{FIG:3}
\end{figure}

 \section{Numerical results and discussions}
\subsection{Effect of surface tension on bubble growth}
In this section, we present the effect of surface tension on the
nonlinear growth rate of the bubble tip for combined Rayleigh--Taylor
and Kelvin--Helmholtz instability. To describe the dynamics of the
bubble tip, it is essential to integrate Eqs. (10)--(12), (15) and
(16) by numerical simulation.  To obtain the initial conditions of
the numerical integration, we assume that the initial interface is
given by $y=\eta_{0}(t=0) cos(k x)$. The expansion of the cosine
function gives $(\xi_{2})_{initial}= 0$ and $(\xi_{3})_{initial}=
-\frac{1}{2}(\xi_{1})_{initial}$, where $(\xi_{1})_{initial}$ is
the arbitrary initial amplitude. Since the perturbation starts
from rest, we may often choose
$(\xi_{4})_{initial}=(\xi_{5})_{initial}=0$. The
non-dimensionalized time development plots of $\xi_1$, $\xi_2$,
$\xi_3$,$\xi_4$ and $\xi_5$ are shown in Figs. 1, 2 and 3.

Before we describe the nature of the bubble tip, consider the
asymptotic behavior of the tip. As $\tau \rightarrow\infty$, the
asymptotic values of $\xi_{3}$, $\xi_{4}$ and $\xi_{5}$ for bubble
are obtained by setting $\frac{d\xi_3}{d\tau}=0$,
$\frac{d\xi_4}{d\tau}=0$ and $\frac{d\xi_5}{d\tau}=0$. Note that,
if $k^2< 3\left(1+\frac{15}{16}\frac{\rho_l}{\rho_h-\rho_l}(\Delta
V)^2 \right)k_c^2$, where $\Delta V=V_{h}-V_{l}$, the asymptotic
values are

\begin{eqnarray}\label{eq:20}
[(\xi_3)_{asymp}]_{bubble}=-\frac{1}{6}
\end{eqnarray}
\begin{align}\label{eq:21}
&[(\xi_4)_{asymp}]_{bubble} \\
&=\sqrt{\frac{2A}{3(1+A)}\left(1-\frac{k^2}{3k_c^2}\right)+\frac{5}{16}\frac{1-A}{1+A}(\Delta
V)^2} \notag
\end{align}
and

\begin{eqnarray}\label{eq:22}
[(\xi_5)_{asymp}]_{bubble}= 0
\end{eqnarray}
where, $A=\frac{\rho_h-\rho_l}{\rho_h+\rho_l}$ is the Atwood
number.

 It is clear form Fig. 1 that surface tension
suppresses the velocity and growth of the bubble tip
significantly, provided surface tension is larger than a critical
threshold, $T<T_c^{bubble}$, where

\begin{eqnarray}\label{eq:23}
T_c^{bubble}= 3\left((\rho_h-\rho_l)+\frac{15}{16}\rho_l(\Delta
V)^2 \right)\frac{g}{k^2}
\end{eqnarray}
Here the critical value depends on the magnitude of relative
velocity shear of two fluids and the growth and velocity of the
tip reduced if $T < T_c^{bubble}$.

When there is no tangential velocity difference (i.e., $V_h=V_l$)
between the two fluids initially, the fluids are purely prone to
the Rayleigh--Taylor instability and the critical value becomes
$\frac{3(\rho_h-\rho_l)g}{k^2}$. These results agree with the
argument in Ref. \cite{ss09}. In absence of surface tension, the
asymptotic values coincide with the results obtained in our
previous work {\cite{rb12}}.

Further, if $T>T_c^{bubble}$ , oscillatory state emerges even
for $r>1$. Figures 2 and 3 describe the oscillatory state of
the motion. The amplitude and  the period of oscillation decrease
monotonically for large surface tension (Fig. 2), while the
amplitude of oscillation increases for large relative velocity
shear (Fig. 3).  In this respect, Figs. 2 and 3 show that
there always exists a self generated oscillatory transverse
velocity component ($-\xi_{5}$) due to perturbation and this
depends upon surface tension as well as the relative velocity
shear $\Delta V$ at the two fluids interface. For negative
velocity shear (i.e, $\Delta V <0$),  the self generated
oscillatory transverse velocity of the bubble peak acts opposite
to the direction of $V_{h}$ and the amplitude of oscillation
increases for large surface tension.

If $T=T_c^{bubble}$, equilibrium is attained, i.e,

\begin{align}\label{eq:24}
&\dot{\xi}_{3}=\dot{\xi}_{4}=\dot{\xi}_{5}=0 \notag \\
&\mbox{ when } \xi_3=-\frac{1}{6} \mbox{ and   }\xi_4=\xi_5=0
\end{align}
and the equilibrium becomes unstable. This feature is shown with a 
dot-dash line in Fig. 1. Thus, the combined Rayleigh--Taylor and
Kelvin--Helmholtz instability is stabilized when

\begin{align}\label{eq:25}
&k^2> 3\left(1+\frac{15}{16}\frac{\rho_l}{\rho_h-\rho_l}(\Delta
V)^2 \right)k_c^2, \notag\\
&\mbox{i.e., } T> T_c^{bubble}
\end{align}
while the instability however persists but with reduced growth
rate for

\begin{align}\label{eq:26}
&k^2\leq 3\left(1+\frac{15}{16}\frac{\rho_l}{\rho_h-\rho_l}(\Delta
V)^2 \right)k_c^2, \notag \\
&\mbox{i.e., } T\leq T_c^{bubble}
\end{align}

According to the condition (25), for $\rho_h=3$, $\rho_l=2$,
$V_h=0.5$ and $V_l=0.1$, the motion is stabilized when
$\frac{k^2}{k_c^2} > 3.9$. These results are exhibited in Fig. 2, where $\frac{k^2}{k_c^2} > 3.9$. For $\frac{k^2}{k_c^2}
=3.9$, the growth rate of the instability is asymptotically
diminished and becomes $0$ (dash-dot line of Fig. 1). However,
Fig. 1 shows the suppression of growth rate of the instability
due to surface tension, when $\frac{k^2}{k_c^2} < 3.9$.

\begin{figure}
%\vspace{-3cm} \vbox{\hskip 1.cm \epsfxsize=15cm \epsfbox{fig4.ps}}
\begin{center}
\includegraphics[width=\columnwidth, trim=0 0cm 0 0, clip]{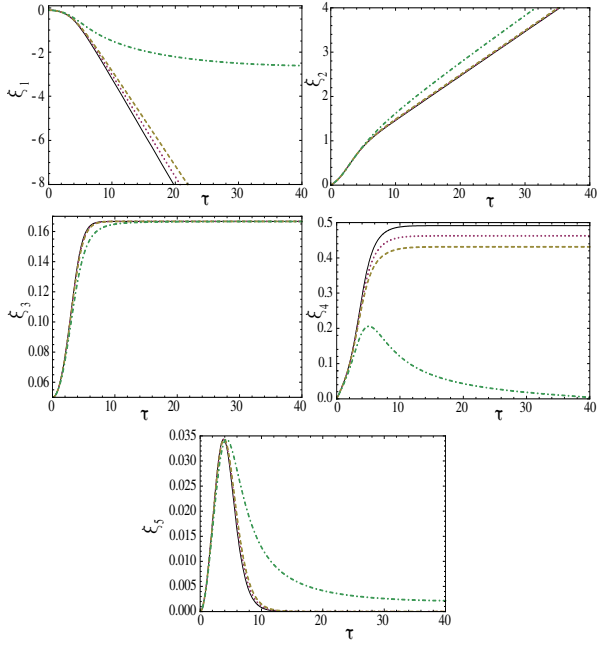} 
\end{center}
\caption{Spike- variation of $\xi_1$, $\xi_2$,
$\xi_3$, $\xi_4$ and $\xi_5$ with $\tau$. Initial value
$\xi_1=$-0.1, $\xi_2=$0, $\xi_3=$0.05, $\xi_4=0$, and $\xi_5=0$
with $\rho_h=3$, $\rho_l=2$, $V_{h}$=0.5, $V_{l}=0.1$,
$\frac{k^2}{k_c^2}$=0 (line), 0.5 (dot), 1 (dash), 4.35 (dash-dot).
} \label{FIG:4}
\end{figure}

\begin{figure}
%\vspace{-5cm} \vbox{ \hskip 1.cm \epsfxsize=15cm \epsfbox{fig5.ps}}
\begin{center}
\includegraphics[width=\columnwidth, trim=0 0cm 0 0, clip]{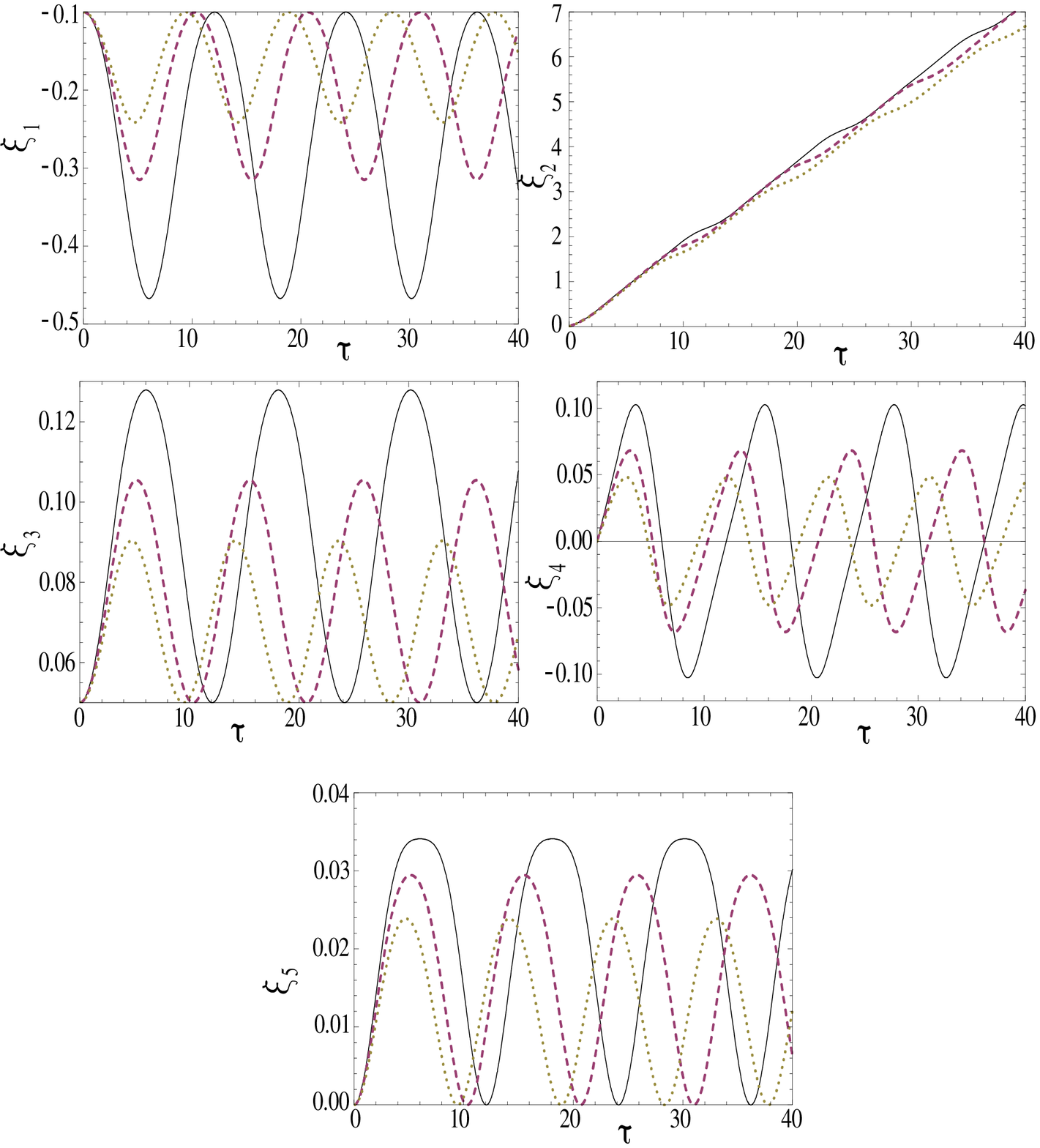} 
\end{center}
\caption{Spike- variation of $\xi_1$, $\xi_2$,
$\xi_3$, $\xi_4$ and $\xi_5$ with $\tau$. Initial value
$\xi_1=$-0.1, $\xi_2=$0, $\xi_3=$0.05, $\xi_4=0$, and $\xi_5=0$
with $\rho_h=3$, $\rho_l=2$, $V_{h}$=0.5, $V_{l}=0.1$,
$\frac{k^2}{k_c^2}$=10 (line), 15 (dot), 20 (dash). } \label{FIG:5}
\end{figure}

\subsection{Effect of surface tension on spike growth}
The temporal evolution of spike state is exhibited in Figs. 4, 5 and 6; the results follow from the numerical integration of
Eqs. (10)--(12), (15) and (16) using the transformation
$\xi_1\rightarrow -\xi_1$, $\xi_3\rightarrow -\xi_3$,
$g\rightarrow -g$, $r\rightarrow\frac{1}{r}$ and
$V_h\rightleftharpoons V_l$. The saturation curvature and velocity
of the spike tip are given by

\begin{eqnarray}\label{eq:27}
[(\xi_3)_{asymp}]_{spike}=\frac{1}{6}
\end{eqnarray}
\begin{align}\label{eq:28}
&[(\xi_4)_{asymp}]_{spike} \\
&=\sqrt{\frac{2A}{3(1-A)}\left(1-\frac{k^2}{3k_c^2}\right)+\frac{5}{16}\frac{1+A}{1-A}(\Delta
V)^2} \notag
\end{align}
and

\begin{eqnarray}\label{eq:29}
[(\xi_5)_{asymp}]_{spike}= 0
\end{eqnarray}
provided $k^2<
3\left(1+\frac{15}{16}\frac{\rho_h}{\rho_h-\rho_l}(\Delta V)^2
\right)k_c^2$.

\begin{figure}
%\vspace{-2cm} \vbox{ \hskip 1.cm \epsfxsize=15cm\epsfbox{fig6.ps}}
\begin{center}
\includegraphics[width=\columnwidth, trim=0 0cm 0 0, clip]{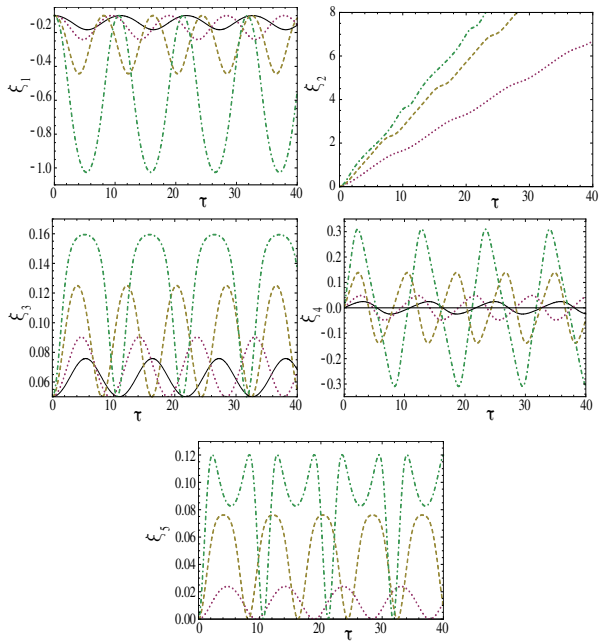} 
\end{center}
\caption{Spike- variation of $\xi_1$, $\xi_2$,
$\xi_3$, $\xi_4$ and $\xi_5$ with $\tau$. Initial value
$\xi_1=$-0.1, $\xi_2=$0, $\xi_3=$0.05, $\xi_4=0$, and $\xi_5=0$
with $\rho_h=3$, $\rho_l=2$, $\frac{k^2}{k_c^2}$=20,
$V_{h}$=0, $V_{l}=0$ (line), $V_{h}$=0.5, $V_{l}=0.1$ (dot),
$V_{h}$=1, $V_{l}=0.1$ (dash), $V_{h}$=1.5, $V_{l}=0.1$ (dash-dot). }
\label{FIG:6}
\end{figure}

Figure 4 describes that large surface tension suppresses the growth rate of the spike tip, as well as the bubble.  The
nonlinear oscillation of the spike tip is observed for $k^2>
3\left(1+\frac{15}{16}\frac{\rho_h}{\rho_h-\rho_l}(\Delta V)^2
\right)k_c^2$ and the equilibrium state arises when equality
holds. The pattern of amplitude and period of oscillation are
identical to that for the bubble (Figs. 5 and 6). Figure 5
shows the oscillatory behavior of the spike structure for
different values of surface tension while the dependency of the
relative velocity shear is demonstrated in Fig. 6.

\section{Conclusion}
In this paper, we have studied a potential flow model  to describe
the nature of the nonlinear structure of a two-fluid interface under the
combined action of Rayleigh--Taylor and Kelvin--Helmholtz
instabilities due to surface tension.  The analytic expressions
for bubble and spike growth rates at asymptotic stage are obtained
for arbitrary Atwood number and velocity shear. Surface tension
becomes a stabilizing factor of the instability, provided it is
larger than a critical value.  In this case, oscillatory behavior
of motion described by numerical integration of governing
equations. The nature of oscillations depends on both surface
tension and relative velocity shear of two fluids. On the other
hand, below the critical value, surface tension dominates the
growth and growth rate of the instability. This result is expected
to improve the understanding of the stabilization factor for the
astrophysical instability.

\begin{acknowledgements}
This work was supported by the University Grant Commission, Government of India under Ref. No. PSW-43/12-13 (ERO).
\end{acknowledgements}

\end{document}